\documentclass[lettersize,journal]{IEEEtran}
\IEEEoverridecommandlockouts

\usepackage{cite}
\usepackage{amsmath,amssymb,amsfonts}
\usepackage{graphicx}
\usepackage{textcomp}
\usepackage{xcolor}
\usepackage{graphicx}%
\usepackage{multirow}%
\usepackage{amsmath,amssymb,amsfonts}%
\usepackage{amsthm}%
\usepackage{mathrsfs}
\usepackage{wrapfig}%
\usepackage{xcolor}%
\usepackage{braket}
\usepackage{textcomp}%
\usepackage{manyfoot}%
\usepackage{booktabs}%
\usepackage{algorithm}%
\usepackage{algorithmicx}%
\usepackage{algpseudocode}%
\usepackage{listings}%

\begin{document}

\title{Towards A Global Quantum Internet: A Review of Challenges Facing Aerial Quantum Networks\\}

\author{%
  Nitin~Jha\textsuperscript{\dag *}, \quad Abhishek~Parakh\textsuperscript{\dag}\\
    \textsuperscript{\dag}Computer Science Department, Kennesaw State University, Marietta, GA 30068\\[1ex]
  \textsuperscript{*}\texttt{njha1@students.kennesaw.edu}\\

}

\maketitle

\begin{abstract}
Quantum networks use principles of quantum physics to create secure communication networks. Moving these networks off the ground using drones, balloons, or satellites could help increase the scalability of these networks. This article reviews how such aerial links work, what makes them difficult to build, and the possible solutions that can be used to overcome these problems. By combining ground stations, aerial relays, and orbiting satellites into one seamless system, we move closer to a practical \textit{quantum internet}. 

\end{abstract}

\begin{IEEEkeywords}
Aerial quantum communication networks, large-scale quantum networks, quantum internet, challenges and solutions.
\end{IEEEkeywords}

\section{Introduction}
In the current age, where the general public has access to significant computational power, keeping information safe from bad actors is paramount. From today's smart refrigerators and toasters to space communications and national defense networks, the need for secure encryption techniques is skyrocketing. With the power of quantum computers on the rise, traditional classical encryption methods may become obsolete within a decade. This introduces the need for a new communication framework that is ``quantum-proof". Unlike classical encryption techniques, which rely on the assumptions of mathematical hardness, quantum communication provides security based on laws of quantum physics. 

Arguably, quantum networks represent the future of secure communication, harnessing quantum phenomena such as \textit{entanglement} and \textit{superposition} to transmit quantum bits (qubits) across long distances. Rather than relaying classical $0$s and $1$s via standard fiber or wireless links, these systems share \textit{entangled} particles or transmit single qubits so that any attempts to compromise the network by bad actors is immediately detectable, ensuring unparalleled security through quantum key distribution (QKD). In the longer term, this approach also promises to interlink remote quantum processors, enabling distributed quantum computing capabilities far beyond today’s classical architectures. 

Large scale quantum networks are currently at a proof-of-concept stage. Teams around the world have built laboratory links and small-scale city networks—such as China’s fiber-optic QKD line between Beijing and Shanghai and Europe’s pilot nodes under the Quantum Internet Alliance—while satellites have begun to relay entangled photons through free space. One key-aspect to developing a true non-terrestrial quantum communication network is developing aerial quantum networks which can be introduced into current ground to satellite quantum networks, which can increase the efficiency and scalability of the integrated quantum network. 

\begin{figure}[h!]
    \centering
    \includegraphics[width=\linewidth]{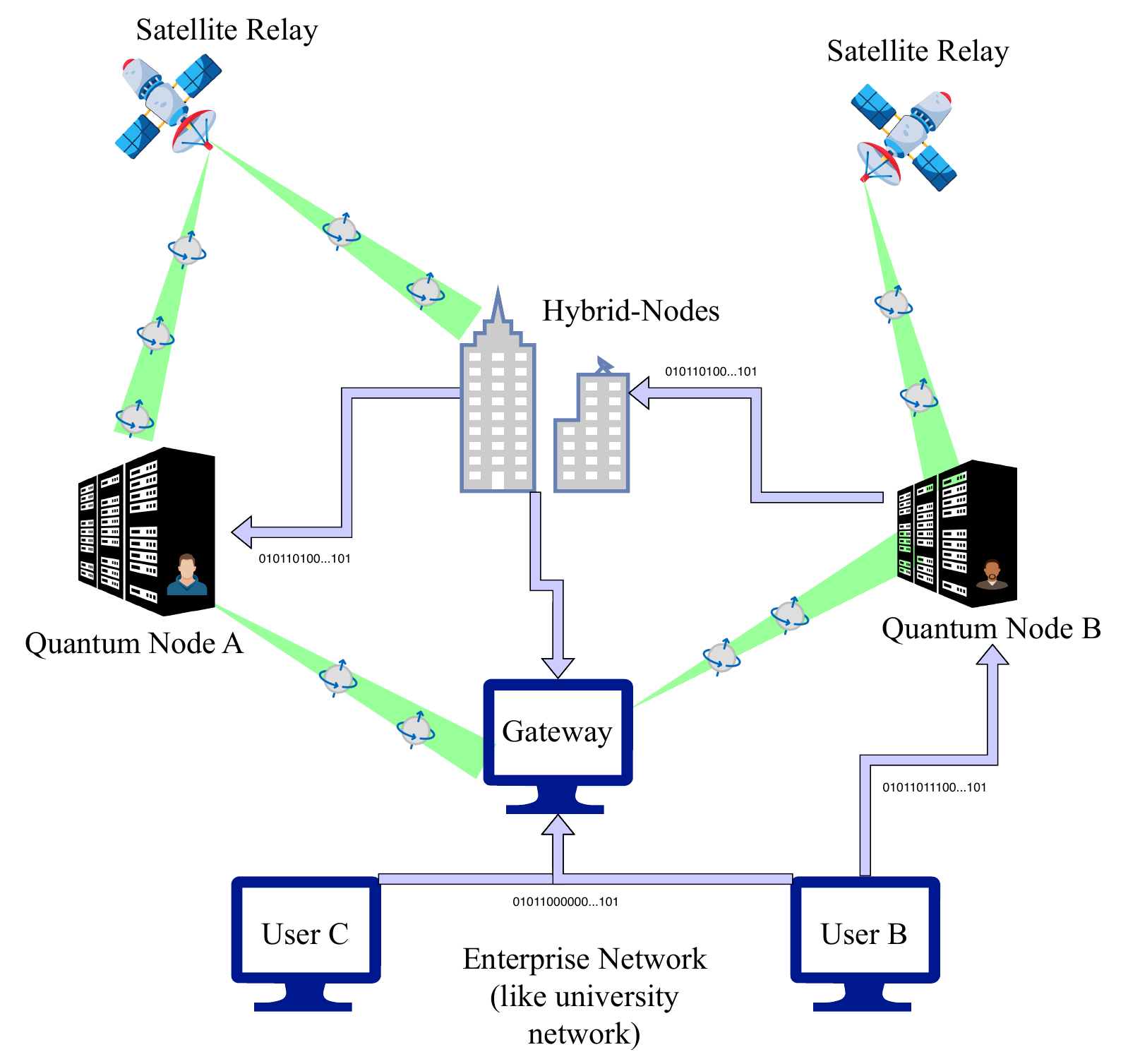}
    \caption{A schematic diagram of quantum network consisting of multiple nodes. Green light cones represents transmittance of qubits, and purple lines denote the flow of classical messages. A quantum network in current stage relies on transmission of both quantum and classical bits. }
    \label{fig:quant-network}
\end{figure}

Figure (\ref{fig:quant-network}) illustrates a typical quantum network. It consists of multiple nodes, some of which can only process quantum information such as \textit{Quantum Node A} and \textit{Quantum Node B}, whereas there are infrastructure which can handle both classical and quantum bits. The \textit{hybrid nodes} in fig(\ref{fig:quant-network}) can process both quantum and classical information and can be relay points for several sub-networks within that group. For example, and enterprise \textit{gateway} can convert the qubits back to classical bits, and thus helps all the other users in the \textit{enterprise network} to work with just classical information to reduce the overall overhead. 

If global quantum communication is to become reality, aerial, free-space and satellite based quantum communication will be the primary means of implementation. In recent times there have been several ground based, and ground-to-satellite quantum key distribution networks. Most famous one is the Chinese satellite experiment where the scientists established entanglement distribution over $1000$km with key-rate useful enough for a metropolitan communication network \cite{liao2017satellite}. Around the same time, high-altitude platforms (HAPS), UAV-based testbeds, and drones have extended free-space quantum links into the stratosphere, proving that mobile aerial relays can carry entangled photons across tens of kilometers under realistic turbulence conditions \cite{conrad2021drone}. Due to the development of smaller, lighter optical devices, like laser sources, it’s now possible to put quantum communication gear on drones without bulky or power-hungry equipment. While geostationary satellites can hover over one spot on Earth, they orbit at about $36,000$ km up—so signals travel farther, become weaker, and take longer to arrive. By combining ground-based fiber links, drone relays, low-Earth-orbit satellites, and geostationary nodes—and adding signal boosters, smart optics, and precise timing—we can build a layered quantum internet that’s truly global, fast, and secure \cite{burrUbiquity2022}.

However, this new framework comes with its challenges. Quantum systems are highly susceptible to environmental disturbances. For example, optical fibers are used for transmitting qubits (the quantum equivalent of bits) in optical-quantum networks. While these fibers work well over short distances, they begin to lose signal over longer distances. As an alternative, LEO (Lower-Earth Orbit) and MEO (Medium-Earth Orbit) satellites can provide global coverage, but each satellite is only overhead for a limited time, thus requiring, large numbers that are extremely costly to deploy \cite{roberts2000lost}. On the other hand, geostationary satellites hover high above the equator, giving constant coverage but at the cost of weak, delayed signals that need large, high-power ground stations and suffer in bad weather. Their crowded orbital ``slots” and limited frequency bands lead to interference and complex coordination. All this makes them expensive to launch and operate, spurring interest in lower-orbit or hybrid networks for faster, more reliable links. To strike a balance between these two approaches, scientists have proposed aerial quantum networks that use drones in the sky to establish quantum communication links.

In this paper, we explain the working of the aerial quantum networks, and review some of the challenges it faces. This review is crucial to understand how aerial quantum networks can enhance the overall working of non-terrestrial quantum networks by acting as relay points in larger networks. Section \ref{Sec:Basic} reviews the basic quantum principles used in such networks. Section \ref{Sec:Aerial} explains the idea of aerial quantum networks, and several components associated with it. Section \ref{Sec:Challenges} explains challenges which are in the way of deploying aerial quantum networks. Section \ref{Sec:Solution} reviews some possible solutions to the challenges. Section \ref{Sec:Conclusion} concludes this article and discusses future directions.

\section{Quantum Computing Basics}
\label{Sec:Basic}
Before we delve deeper into the nature of aerial quantum networks and the challenges they present, it's important to understand the fundamental principles behind these networks. Simply put, a quantum computers and networks leverage principles of quantum physics to perform the tasks of computing and communications more efficiently and securely, respectively. Quantum networks can act as the \textit{link} that enable secure communication between classical or quantum devices. The basic information unit for a quantum computer is a qubit, the quantum equivalent of a classical bit. Qubits are of several different types, but at the base level, they are inherently subatomic units such as photons, squeezed atoms, or quantum dots.  The following are some of the quantum principles used in computing and communication, 
\begin{enumerate}
    \item \textbf{No-Cloning Principle}: Classical information can be easily duplicated, even if it's encrypted. However, this cannot be done for unknown quantum information. Consequently, quantum computers have to perform ``blind" operations (i.e. allow the quantum states to evolve overtime) before making a measurement. Fig(\ref{fig:no-cloning}) shows a schematic diagram representing the concept of no-cloning principle.

    \begin{figure}[h!]
        \centering
        \includegraphics[width=\linewidth]{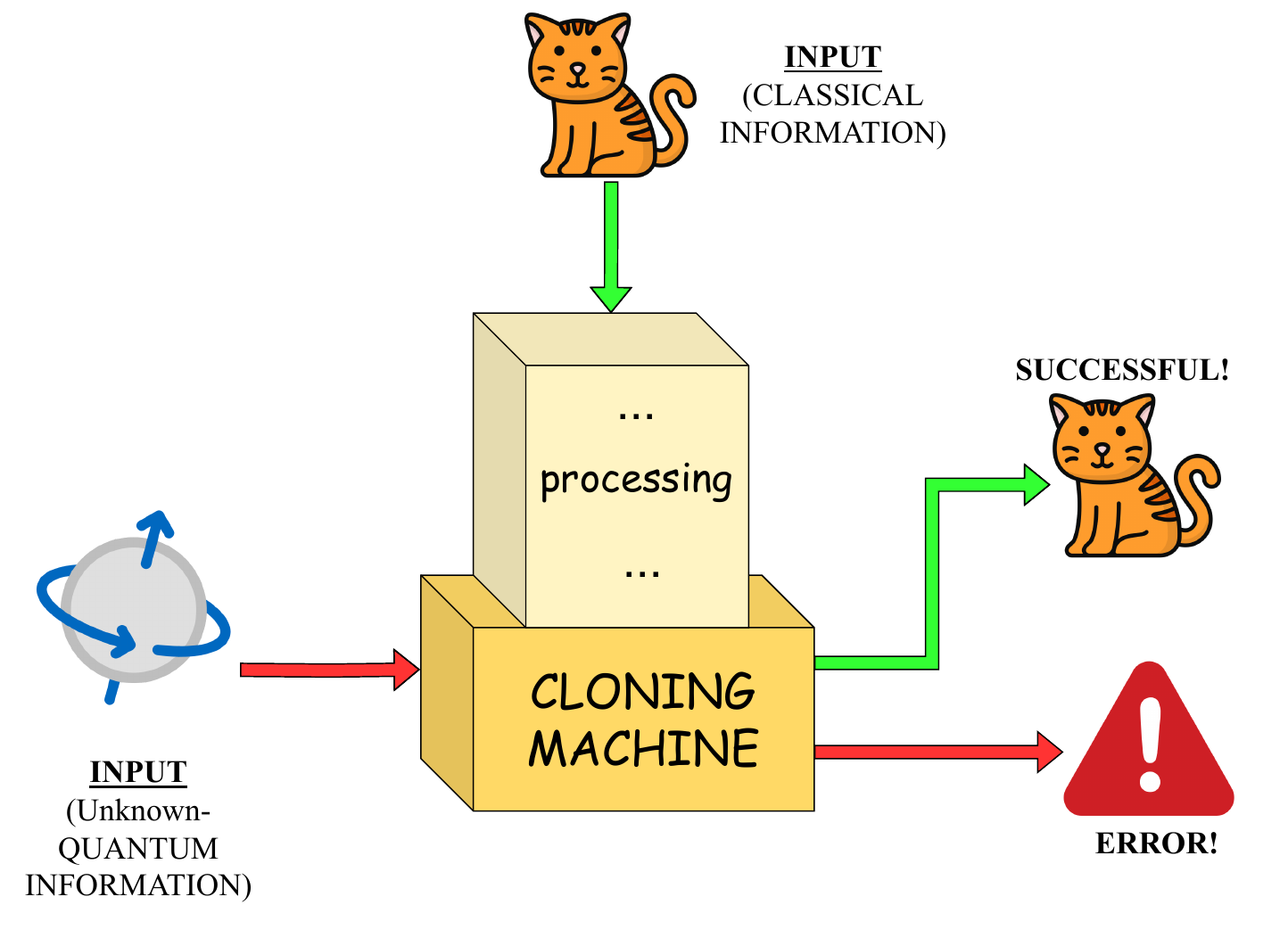}
        \caption{A visual representation of no-cloning principle. }
        \label{fig:no-cloning}
    \end{figure}

    \item \textbf{Quantum Superposition Principle}: A classical states are deterministic. However, quantum information may exist in a \textit{superposition} of states. This helps quantum computers perform calculations faster by evaluating multiple possible combinations at once. This is known as \textit{quantum parallelism}. Moreover, a quantum state exists in a superposition until it's observed, and upon observing it, it collapses to one of those possible states \cite{markidis2024quantum}. Fig(\ref{fig:superposition}) describes the concept of superposition as the famous example of Schr\"{o}dinger's cat. 

    \begin{figure}[h!]
        \centering
        \includegraphics[width=0.85\linewidth]{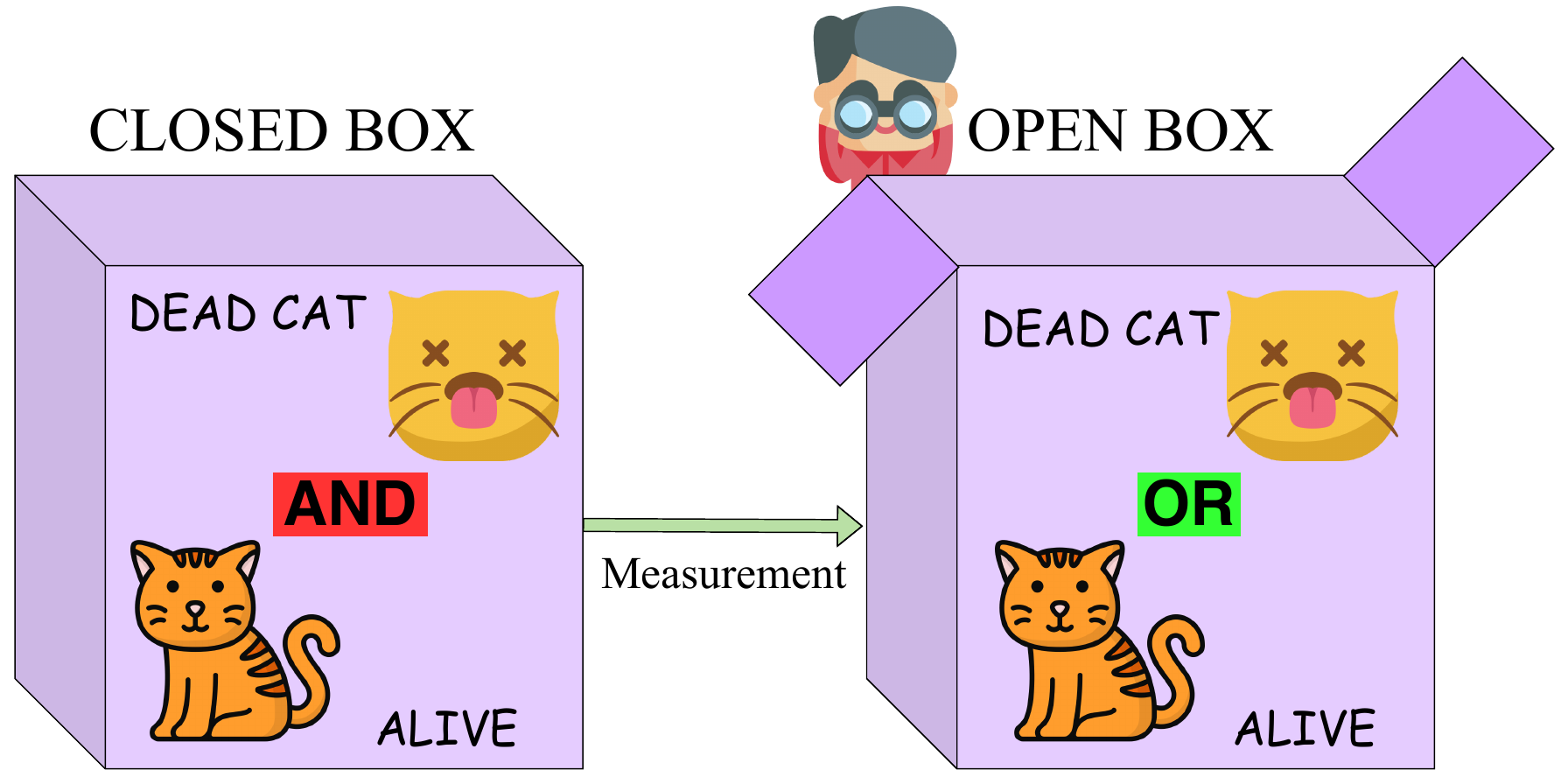}
        \caption{A visual representation of quantum superposition. The Schr\"{o}dinger's cat is both alive and dead (in a superposition) unless we open the box and find it either to be alive, or dead.}
        \label{fig:superposition}
    \end{figure}

    \item \textbf{Quantum Entanglement}: This has been considered one of the strangest physical phenomena where two quantum states are intertwined in such a way that they cannot be distinguished. A measurement on one of these states can predict the state of the other. This was first theorized by Einstein, Podolsky, and Rosen, and thus, one of the most famous entangled states is known as an \textit{EPR} pair. \cite{einstein1935can}

\end{enumerate}

Quantum algorithms typically make use of one or more of the above quantum properties to compute solutions of problems efficiently that are often intractable using classical algorithms or an efficient solution is not know. Similarly, quantum networks rely on these principles to provide security, detect eavesdropping and exchange encryption keys.

\section{Aerial Quantum Communication}
\label{Sec:Aerial}
\begin{figure}[h!]
    \centering
    \includegraphics[width=\linewidth]{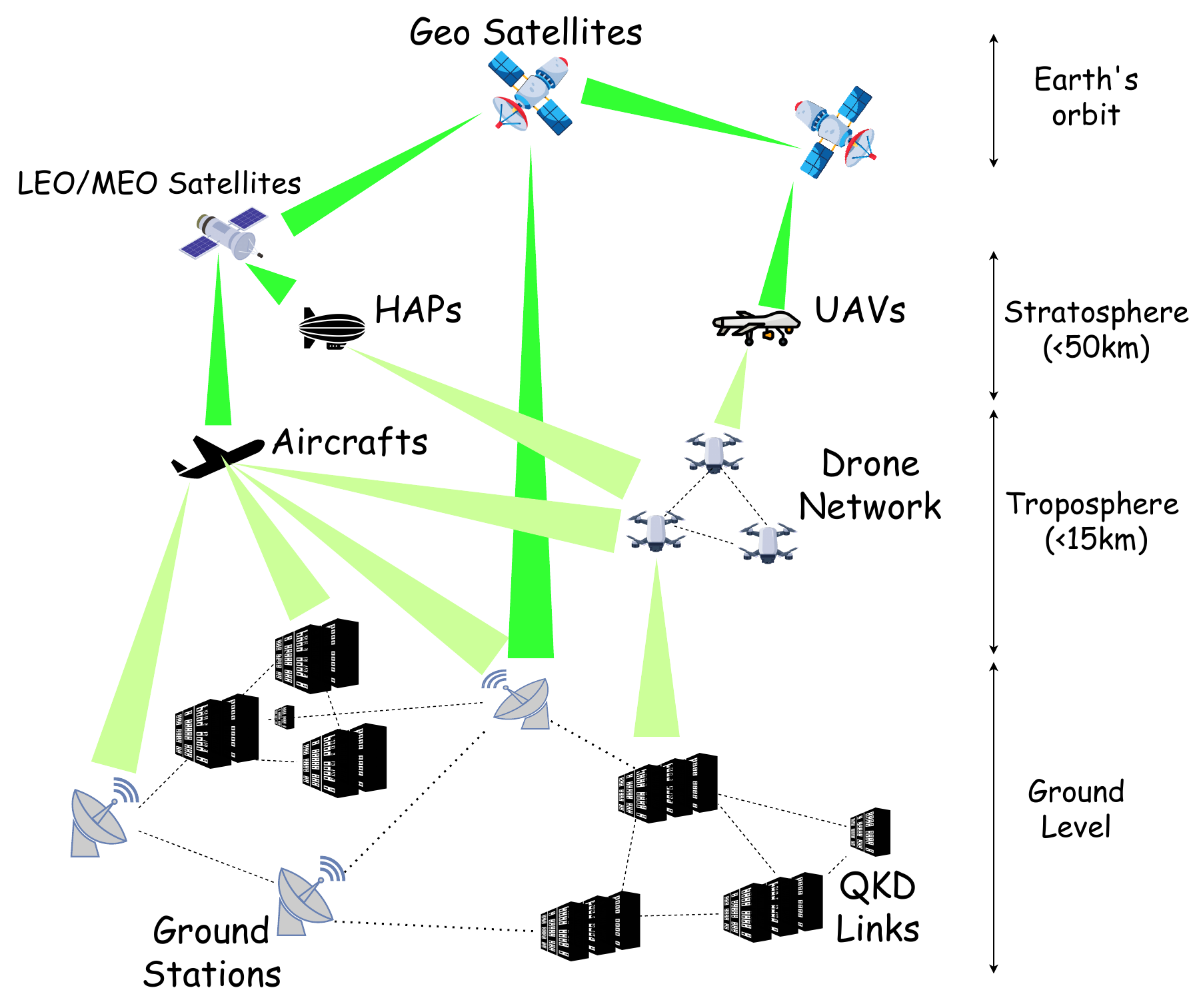}
    \caption{Aerial-Quantum Communication Network Diagram. There are several layers in this network setting. There are ground based QKD networks, the lower-altitude drone-network, inter-connected network between aircraft, HAPs, and UAVs, and finally the satellite to ground based communication network. This diagram illustrates how non-terrestrial quantum networks work alongside currently existing satellite to ground based networks.}
    \label{fig:aerial}
\end{figure}

Aerial quantum communication consists of using aerial platforms such as drones, hot air-balloons, aircraft, etc. to establish secure communication.  In an aerial quantum network, ground stations send quantum signals upward to the aforementioned aerial platforms, which then act as temporary relays, passing the quantum information along until it reaches its final destination. Deploying such a network offers several benefits such as setting up in disaster zones, difficult topographical zones where building permanent ground stations is not feasible. Fig(\ref{fig:aerial}) provides a schematic diagram about how different quantum links (green cones show open-air links, and dotted lines show physical cable links on the ground) can be established between different levels: ground, aerial, and space.

As useful as they are, deploying such networks have several physical challenges such as sensitivity of quantum signals to atmospheric conditions, unpredictability of atmospheric conditions, slight motion due to unseen weather can disrupt necessary alignments. There are major advancements in optical tracking systems and lightweight quantum equipment are making it possible to imagine a future where flying drones and balloons routinely carry quantum information across cities, deserts, and oceans, creating secure, mobile communication networks. 

\section{Practical Challenges}
\label{Sec:Challenges}
As discussed above, using aerial platforms as quantum relay nodes to develop an aerial quantum network comes with several technological challenges from atmospheric uncertainty to practical limitations. 

\subsection{Atmospheric Turbulence}
\begin{figure}[h!]
    \centering
    \includegraphics[width=\linewidth, alt={Aerial communication network with air turbulence profile}]{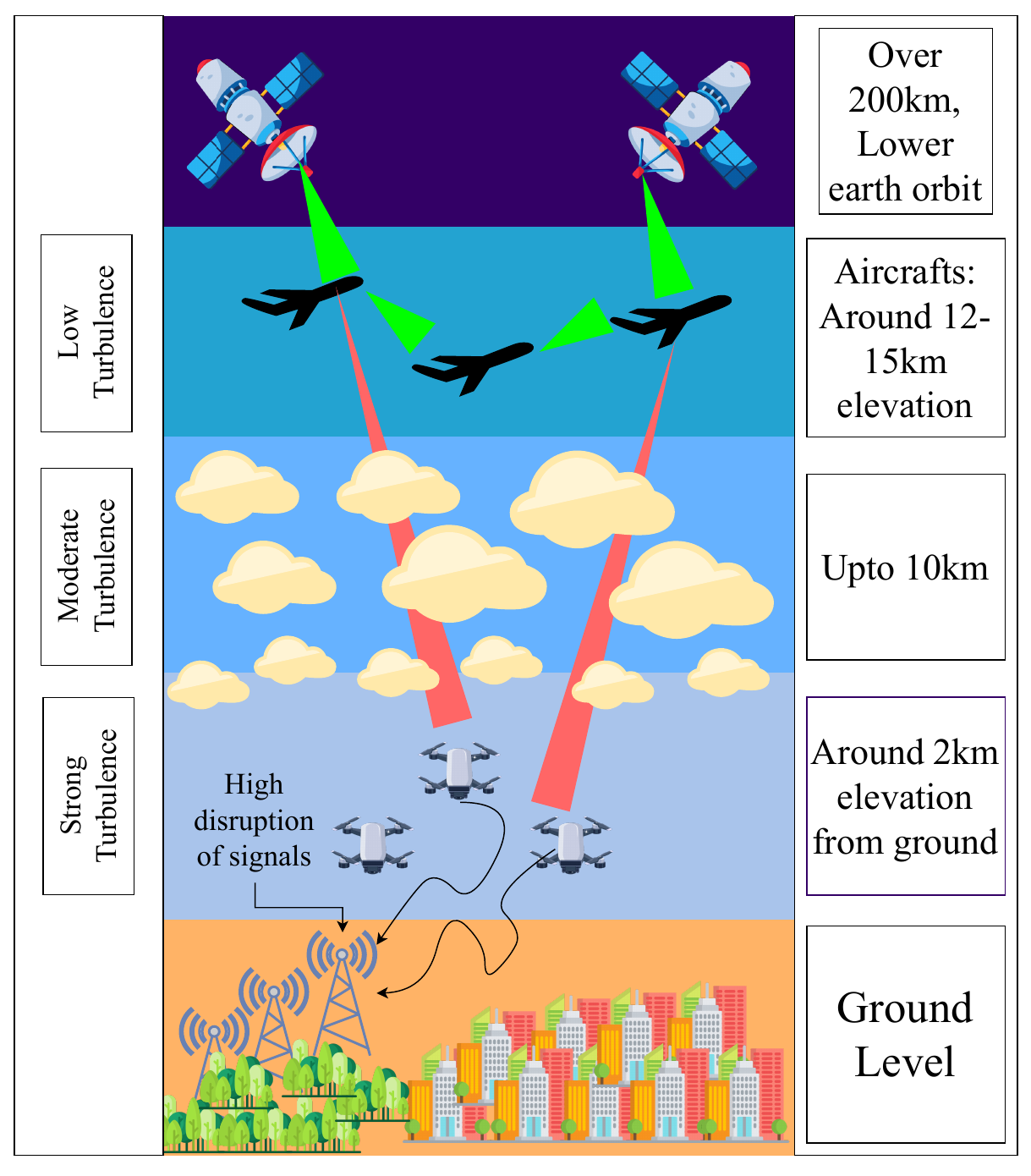}
    \caption{Different components of aerial communication network with air turbulence profile. The turbulence at different altitudes are shown on the left axis, where we can see drones operate in severely high turbulent zones and satellites operates in the least turbulent zones. The main idea behind using drones in these network setting is to provide relay stations for satellite to ground based communication, such that the distance of travel for qubits is also reduced by use of these relay points in different atmospheric levels (aircraft and then drones), thus attaining a better quality of transmission. }
    \label{fig:turbulence}
\end{figure}

Air turbulence around the surrounding atmosphere is one of the most significant factors for \textit{free-space optical networks}, i.e., networks where optical signals are sent across air without any use of \textit{physical} cables. So, when we are trying to send some qubits, in the form of laser beams, across the ``free-space" the information will not always go in a straight path. We know that the atmosphere is ever changing-- the pressure, temperature, and density shifts-- that leads to air turbulence. Aerial quantum communication uses drones which majorly operates in very turbulent zones as also represented in fig(\ref{fig:turbulence}).

The turbulence not only affects air-vehicles, but also effects these traveling laser beams in three major ways: (1) \textbf{bends the wave}, i.e., changes its path, (2), \textbf{scatters the wave} and (3) \textbf{weakens the intensity} of these laser pulses \cite{laserna2009study}. The main causes behind these three effects can be explained by the following phenomena,

\begin{enumerate}
    \item \textbf{Scintillation and Beam Wandering}: The constant changes in the air conditions causes the signal pulses to bend and distort. This results in the receiver receiving \textit{flickering} or \textit{speckled} patterns, which is known as scintillation. Scintillation results in random high and low intensity jumps in the signal, which results in an unexpected signal-to-noise ratio, which in turn makes the communication unreliable. One way to solve this is by using a larger aperture at receiver end to average out these noise and effectively smoothen out the noisy signal \cite{churnside1991aperture}.
    
    \textit{Beam wandering} refers to the deviation of the signal from its intended path. This happens mainly because of two reasons (1) atmospheric vulnerabilities (i.e., the ways in which the air itself can disturb a tightly focused light beam as it travels) and (2)  misalignment of transmitters that causes the wave to wobble (like small mechanical jitters).  Recent works identify following three issues, 
    \begin{itemize}
        \item Beam spread dominates for longer distance traveled.
        \item Wandering due to turbulence (small yet significant).
        \item Pointing errors are small but significant especially at shorter distances.
    \end{itemize}
    At nighttime, the air condition tends to be more stable around the ground, however small pockets of turbulence still exist. These tiny swirls in the air can slightly bend and distort laser beams as they travel. Even though nighttime generally offers better conditions for quantum communication compared to the daytime, these small-scale disturbances still cause some beam wandering and spreading.

    \item \textbf{Atmospheric Attenuation}: Even if the sky seems clear, its never perfectly transparent. So, whenever a quantum signal is transmitted through open-air, there's water-vapor, fog, sometimes rain, etc., which effectively weakens or scatters the signal. This gradually causes the signal to weaken as it passes through the atmosphere, and this effect is known as \textit{atmospheric attenuation}. This results in the loss of energy of the quantum signals, which typically happens in the following ways, 
    \begin{itemize}
        \item \textbf{Absorption} by water vapor or other molecules in the air. 
        \item \textbf{Scattering} of quantum signals by these molecules in the air (such as dust, water vapor). 
        \item \textbf{Scintillation} as described above (repeat briefly).
    \end{itemize}
    It's quite obvious at this point that the adverse weather conditions result in more atmospheric attenuation since there would be more water vapor, fog, etc., in the air. Studies show that in very thick fog or heavy clouds, all wavelengths of light are blocked about the same, but in light fog or haze, deeper-red light—like infrared—passes through much more easily \cite{dubey2024review,korevaar1999optical}.

    \item \textbf{Beam Divergence Loss}: A light beam when traveling across free space often spreads out rather than traveling as a sharp beam of light. This is referred to as \textbf{beam divergence}. As the beam diverges more and more, only a smaller portion of light would be captured by the receiver. Thus, losing some of the signal. This loss is called \textbf{beam divergence loss} and it is mainly dependent on the following parameters,
    \begin{itemize}
        \item The original beam size at the transmitter (sender end). 
        \item Distance the beam has to travel.
        \item Wavelength of the optical signal used. 
    \end{itemize}
    This problem can be tackled by adjusting the aperture of the transmitting and receiving optics. A bigger transmitter can send a targeted beam, and a bigger receiver can catch more of the diverged beam, however bigger optics usually come at a great tradeoff between cost and efficiency.
\end{enumerate}

\section{Possible Solutions}
\label{Sec:Solution}

As discussed in previous sections, there are several atmospheric challenges to deploying an aerial quantum network. However, several recent studies have worked on finding possible resolutions to make these aerial quantum networks more scalable in context of the quantum internet. In this section, we would briefly go over two of the key-solutions highlighted in some recent works. 

\subsection{Hybrid Model}
The major implementation issue for aerial quantum networks are due to atmospheric instabilities. The core difficulty lies in accurately quantifying how much of the quantum signal, subject to many variable factors, actually reaches the receiver. To address these issues, several studies have developed a model for low-altitude aerial quantum networks. In this model, the laser beam is treated not as a perfect point but as an elliptical ``blob” whose position and shape fluctuate randomly during flight. It explicitly accounts for both atmospheric turbulence—which distorts the beam—and pointing errors introduced by the drone or ground station.

\subsection{Regarding Network Links}
Aerial networks consists of quantum signal being transmitted twice over the open-air, once from ground station up to the aerial platform (\textbf{up-link}), and then from the aerial platform back to the ground station (\textbf{down-link}). The turbulence near the ground majorly affects the up link, whereas pointing accuracy affects the down-link. 

\begin{enumerate}
    \item \textbf{Link Configuration}: The first thing is setting up reliable quantum links between ground stations and aerial platforms. While the beams are on up-link, they are severely affected by strong turbulence near the ground level which causes the beam to diverge, and other different losses discussed earlier. While sending the beams on down-link, the major issue is pointing alignment. However, for aerial setups like drones the difference between up-link and down-link are not huge. For short distances, both types of links experience only minor deviations in terms of atmospheric losses and beam spread. That means drones can be flexible: either sending or receiving quantum signals without needing radically different equipment designs.
    \item \textbf{Link Budgeting}: As discussed earlier, when quantum signals travel between ground stations and drones, they lose energy, i.e., intensity. Some key reasons for this are (a) distance signal has to travel, (b) atmospheric conditions, (c) beam divergence loss, (d) addition of background noises, and (e) optical loses. Therefore, before setting up secure communication we need to compare the signal strength to incorporate all of these losses. This is called link budgeting. 
    \item \textbf{Link Margin}: Now that we have link budgeting in place, we need to adjust signals such that the receiver gets appropriate intensity of the signal inclusive of all the loses discussed. Link margin is the extra signal strength that we have over the defined link budget. There two cases of link margin, 
    \begin{itemize}
        \item Positive link margin means a reliable link even with a little worse weather conditions.
        \item Negative link margin means the link is susceptible to fail with worsening weather conditions. 
        
    \end{itemize}
    Using larger apertures for receiver can boost the link margin, helping the system stay operational over longer distances. 
\end{enumerate}

\subsection{Time Synchronization}
In quantum communication, aligning the receiver and sender optical equipments are not enough. We also need to make sure to align the timing of sent and received signals too. With aerial quantum communication network using drone, this becomes harder than ground based networks. As the distance between sender and receiver changes constantly, synchronizing their clocks becomes a dynamic, real-time problem. This problem can be solved using several ways such as the use of stable reference clocks, compensation for any delays in signals, and other error correction methods. In fact, quantum clock synchronization experiments have already achieved picosecond-level precision — meaning they can align clocks to within a trillionth of a second.

We have reviewed the current state of aerial-quantum networks, such as use of drones and UAVs or HAPs to act for relay nodes for qubits transmission. We also looked over several practical challenges associated with it, the most important being the problem of atmospheric turbulence. We also looked over possible solutions to these problems such as use of low-altitude networks, and adjusting the several network links to tackle different atmospheric losses. The road-map for such aerial quantum network has many obstacles, however, the on-going research brings us ever closer to a more practical aerial quantum network.

\section{Conclusion}
\label{Sec:Conclusion}
Over the years, there have been proposals for deployment of non-terrestrial quantum communication networks to improve quantum key-distribution standards, such as stable key-rates over longer distances. There have been joint and individual efforts by government agencies, industry, military, and academia to advance the field of quantum communication from ground based to combination of ground and aerial based quantum communication network. The current issues surrounding both ground-based or satellite-based quantum communication network can be solved by aerial quantum networks. This approach, however, comes with its own challenges such as atmospheric vulnerabilities, alignment issues, need for smaller optical devices, and synchronization issues. 

Looking ahead, a layered quantum-internet—unifying terrestrial fibers, aerial platforms, LEO satellite constellations, and GEO nodes—augmented by quantum repeaters, real-time channel estimation, and network-level routing protocols, offers the most promising path to global, low-latency, quantum-secure communications. Achieving this vision will demand close collaboration among optical engineers, quantum-hardware developers, and network architects, along with large-scale field trials and emerging standards. By focusing on the practical hurdles we can address now, the community can build momentum toward a resilient quantum network—one that not only secures today’s data but also supports the high-throughput, distributed quantum computers of the future.

\bibliographystyle{unsrt}
\bibliography{main}

\begin{thebibliography}{10}

\bibitem{liao2017satellite}
Sheng-Kai Liao, Wen-Qi Cai, Wei-Yue Liu, Liang Zhang, Yang Li, Ji-Gang Ren, Juan Yin, Qi~Shen, Yuan Cao, Zheng-Ping Li, et~al.
\newblock Satellite-to-ground quantum key distribution.
\newblock {\em Nature}, 549(7670):43--47, 2017.

\bibitem{conrad2021drone}
Andrew Conrad, Samantha Isaac, Roderick Cochran, Daniel Sanchez-Rosales, Brian Wilens, Akash Gutha, Tahereh Rezaei, Daniel~J Gauthier, and Paul Kwiat.
\newblock Drone-based quantum key distribution (qkd).
\newblock In {\em Free-space laser communications XXXIII}, volume 11678, pages 177--184. SPIE, 2021.

\bibitem{burrUbiquity2022}
Joseph Burr, Abhishek Parakh, and Mahadevan Subramaniam.
\newblock Quantum internet.
\newblock {\em Ubiquity}, 2022(August), August 2022.

\bibitem{roberts2000lost}
Lawrence~D Roberts.
\newblock A lost connection: Geostationary satellite networks and the international telecommunication union.
\newblock {\em Berk. Tech. LJ}, 15:1095, 2000.

\bibitem{markidis2024quantum}
Stefano Markidis.
\newblock What is quantum parallelism, anyhow?
\newblock In {\em ISC High Performance 2024 Research Paper Proceedings (39th International Conference)}, pages 1--12. Prometeus GmbH, 2024.

\bibitem{einstein1935can}
Albert Einstein, Boris Podolsky, and Nathan Rosen.
\newblock Can quantum-mechanical description of physical reality be considered complete?
\newblock {\em Physical review}, 47(10):777, 1935.

\bibitem{laserna2009study}
JJ~Laserna, R~Fern{\'a}ndez Reyes, R~Gonz{\'a}lez, L~Tobaria, and P~Lucena.
\newblock Study on the effect of beam propagation through atmospheric turbulence on standoff nanosecond laser induced breakdown spectroscopy measurements.
\newblock {\em Optics express}, 17(12):10265--10276, 2009.

\bibitem{churnside1991aperture}
James~H Churnside.
\newblock Aperture averaging of optical scintillations in the turbulent atmosphere.
\newblock {\em Applied Optics}, 30(15):1982--1994, 1991.

\bibitem{dubey2024review}
Umang Dubey, Prathamesh Bhole, Arindam Dutta, Dibya~Prakash Behera, Vethonulu Losu, Guru Satya~Dattatreya Pandeeti, Abhir~Raj Metkar, Anindita Banerjee, and Anirban Pathak.
\newblock A review on practical challenges of aerial quantum communication.
\newblock {\em Physics Open}, page 100210, 2024.

\bibitem{korevaar1999optical}
Eric~J Korevaar.
\newblock Optical wireless communications ii.
\newblock Society of Photo-Optical Instrumentation Engineers (SPIE), 1999.

\end{thebibliography}

\end{document}